\documentclass[12pt]{article}
 
\usepackage{epsfig}

\title{Symmetric and asymmetric nuclear matter \\
        in the  Thomas-Fermi model \\
        at finite temperatures\footnote{This investigation
        is dedicated to Georg S\"u{\ss}mann at the occasion
        of his seventieth birthday.}}

\author{K. Strobel, F. Weber, and M. K. Weigel \\
  Sektion Physik, Universit\"at M\"unchen\\ 
  Am Coulombwall 1, D-85748 Garching, Germany 
}
\begin{document}
\baselineskip15pt
\maketitle
%%%%%%%%%%%%%%%%%%%%%%%%%%%%%%%%%%%%%%%%%%%%%%%%%%%%%%%%%%%%%%%%%%%%%%%%%
\begin{abstract}
The properties of warm symmetric and asymmetric nuclear matter
are investigated in the frame of the Thomas-Fermi approximation using
a recent modern parametrization of the effective nucleon-nucleon
interaction of Myers and \'Swi\c atecki.
Special attention is paid to the liquid-gas phase transition,
which is of special interest in modern nuclear physics.
We have determined the critical temperature, critical density
and the so-called flash temperature.
Furthermore the equation of state for cold neutron star
matter was calculated.
\end{abstract}


\newpage
%%%%%%%%%%%%%%%%%%%%%%%%%%%%%%%%%%%%%%%%%%%%%%%%%%%%%%%%%%%%%%%%%%%%%%%%%
\section{Introduction}

An essential but also a very complex and complicated problem of
modern physics is the structure of matter under extreme conditions 
of temperature and/or density. 
The equation of state of hot dense matter is a key ingredient for 
many branches of modern physics.
To mention in this respect are, for instance, the understanding of 
the physics of the early universe, supernova explosions and protoneutron 
stars , respectively.
A further important example are high-energy heavy-ion collisions,
which provide, at present, the only laboratory means to study the behaviour 
of hot and dense matter.
In the literature one can roughly distinguish between two theoretical 
attempts to investigate the equation of state, namly the non-relativistic
and relativistic approaches.
The majority of the relativistic treatments are performed in the 
framework of the relativistic Hartree approximation and only
a few investigations use the relativistic Hartree-Fock or 
relativistic Br\"uckner-Hartree-Fock approximation 
(see, for instance, Refs. \cite{Ser92}-\cite{HWW98}).
The majority of the theoretical treatments utilize the non-relativistic
scheme, using either effective density dependent interactions
or the Br\"uckner approach \cite{BGLBF95}-\cite{DTS92}.

In this investigation we are going to extend a recent modern
Thomas-Fermi approach of Myers and 
\'Swi\c atecki \cite{MS90}-\cite{MS98} to the 
equation of state at finite temperatures (see also Ref. \cite{Str96}).
Furthermore we will briefly discuss the question of neutron
star matter and the equation of state for higher asymmetries,
respectively.

The paper is organized as follows:
First, we recapitulate fore the sake of completeness in 
Section~\ref{sec2} the basic structure of the Thomas-Fermi model
\cite{Tho26, Fer28} for density and momentum dependent interactions.
The extension of the model to finite temperatures is given
in Section~\ref{sec3}.
Section~\ref{sec4} deals with the results of the model and
the last Section is a summary of the results with conclusions.

%%%%%%%%%%%%%%%%%%%%%%%%%%%%%%%%%%%%%%%%%%%%%%%%%%%%%%%%%%%%%%%%%%%%%%%%%
\section{Thomas-Fermi model at $T = 0$} \label{sec2}

We shall not recapitulate the basic assumptions and limitations of the 
Thomas-Fermi model in nuclear physics, since they are described in
greater detail in the pioneering work of Myers and 
\'Swi\c atecki \cite{MS69}.
In the modern version they use an effective interaction of the following 
structure ($\tau$ denotes the isospin):
%%%%%%%%%%%%%%%%%%
\begin{eqnarray}
\label{I.1}
v_{12\tau} &= & - \frac{2T_\mathrm{F,\tau}}{\rho_0}
\,g\,\left(\frac{r_{12}}{a}\right) \\ \nonumber
 && \times \left(\frac{1}{2}(1 \mp \xi) \alpha - \frac{1}{2}(1 \mp \zeta)
\left(\beta \left(\frac{p_{12}}{p_\mathrm{F}}\right)^2 - \gamma \frac{p_\mathrm{F}}{|p_{12}|}
+ \sigma \left(\frac{2 \bar \rho}{\rho_0}\right)^{\frac{2}{3}}\right)\right)~,
\end{eqnarray} 
%%%%%%%%%%%%%%%%%%
with the following definition:
%%%%%%%%%%%%%%%%%%
\begin{equation}
\bar{\rho}^\frac{2}{3} = \frac{1}{2} \left( \rho_{1}^\frac{2}{3} + 
                                          \rho_{2}^\frac{2}{3} \right),
\end{equation}
%%%%%%%%%%%%%%%%%%
here, $\rho_1$, $\rho_2$ denote the (proton or neutron) density at
the locations $\vec r_1$ and $\vec r_2$, respectively.
The density at equilibrium is 
$\rho_0 = (\frac{4\pi}{3} r_0^3)^{-1}$, and the Fermi momentum
and the Fermi kinetic energies are given by 
[$\bar m := \frac{1}{2} (m_n + m_p)$]:
%%%%%%%%%%%%%%%%%%
\begin{equation}
p_\mathrm{F} = \hbar \left( \frac{3}{2} \pi^2 \rho_0 \right)^\frac{1}{3},
\end{equation}
%%%%%%%%%%%%%%%%%%
\begin{equation}
T_\mathrm{F} = \frac{p_\mathrm{F}^2}{2 \bar m},
\end{equation}
%%%%%%%%%%%%%%%%%%
\begin{equation}
T_\mathrm{F,\tau} = \frac{p_\mathrm{F}^2}{2 m_\tau}.
\end{equation}
%%%%%%%%%%%%%%%%%%
By means of the parameters $\xi$ and $\zeta$ the interaction distinguishes 
the forces between nucleons of the same kind (upper sign) and of 
different kind (lower sign).
With the parameter $\alpha$ one adjusts mainly the binding properties
of nuclear matter.
The repulsion is described by the momentum dependent term
$\propto \beta p_{12}^2$.
Since this repulsion turns out to be too strong for higher relative
momenta, one corrects this deficit by the term 
$\propto \gamma |p_{12}|^{-1}$.
In order to obtain better values for higher asymmetries, the 
authors have chosen $\xi \not= \zeta$ in contrast to former 
structures of Seyler and Blanchard \cite{SB61}.
The term proportional to $\sigma (2\bar\rho/\rho_0)^{2/3}$
takes care for a better agreement with the nuclear optical potential
(for more details, see Refs. \cite{MS90}-\cite{Str96}).
Fore our task the model contains six free parameters (in the 
expression for the energy the parameters $\beta$ and $\sigma$
enter only in the combination $\mathrm{B} = \beta + \frac{5}{6} \sigma$)
and they are given in Table~I.
For the space dependence the standard Yukawa form is employed, i.e.:
%%%%%%%%%%%%%%%%%%%
\begin{equation}
   g(\frac{r_{12}}{a}) = \frac{1}{4 \pi a^{3}} 
   \frac{\exp{(-\frac{r_{12}}{a}})}{\frac{r_{12}}{a}} .
\end{equation}
%%%%%%%%%%%%%%%%%%%
The energy per nucleon in the Thomas-Fermi model is given by:
%%%%%%%%%%%%%%%%%%%
\begin{equation}
   u = \frac{E}{N} = \frac{1}{\rho} \frac{2}{(2 \pi \hbar)^{3}}
   \sum_{\tau} \int \limits_{\tau}d^{3}p_{1} \left( \frac{p_{1}^{2}}{2 m_{\tau}} 
   + \frac{1}{2} V_{\tau}(p_{1}) \right) ,
\end{equation}
%%%%%%%%%%%%%%%%%%%
with the single-particle potential:
%%%%%%%%%%%%%%%%%%%
\begin{equation}
   V_{\tau}(p_{1}) = - \frac{2}{(2 \pi \hbar)^{3}}
   \left( \int \limits_{\tau} d^{3}p_{2} v_{12\tau}
   + \int \limits_{-\tau} d^{3}p_{2} v_{12} \right) ,
\end{equation}
%%%%%%%%%%%%%%%%%%%
or more explicitly:
%%%%%%%%%%%%%%%%%%%
\begin{eqnarray}
   V_{\tau}(p_{1}) & = & - \frac{2}{(2 \pi \hbar)^{3}} 
   \frac{2}{\rho_{0}}
   \nonumber\\ 
    & & \times \left[ T_{\mathrm{F}, \tau} \int \limits_{\tau} d^{3}p_{2} \left( 
    \alpha_{l} - \beta_{l} \left( \frac{p_{12}}{p_\mathrm{F}} \right)^{2} 
   + \gamma_{l} \frac{p_\mathrm{F}}{|p_{12}|} 
   - \sigma_{l} \left( \frac{2 \overline{\rho}}{\rho_{0}} \right)^{\frac{2}{3}} 
   \right) \right. 
   \nonumber\\
    & &  \left. + T_\mathrm{F} \int \limits_{-\tau} d^{3}p_{2} \left( \alpha_{u} 
   - \beta_{u} \left( \frac{p_{12}}{p_\mathrm{F}} \right)^{2} 
   + \gamma_{u} \frac{p_\mathrm{F}}{|p_{12}|} 
   - \sigma_{u} \left( \frac{2 \overline{\rho}}{\rho_{0}} \right)^{\frac{2}{3}}
   \right) \right] .
   \nonumber\\ 
    & &
\end{eqnarray}
%%%%%%%%%%%%%%%%%%%
Here, $\alpha_\mathrm{l}$, $\beta_\mathrm{l}$, $\gamma_\mathrm{l}$,
$\sigma_\mathrm{l}$ and 
$\alpha_\mathrm{u}$, $\beta_\mathrm{u}$, $\gamma_\mathrm{u}$,
$\sigma_\mathrm{u}$ are defined as:
$\alpha_\mathrm{l} = 0.5 (1 - \xi) \alpha$; $\beta_\mathrm{l} = 
0.5 (1 - \zeta) \beta$; $\gamma_\mathrm{l} = 0.5 (1 - \zeta) \gamma$;
$\sigma_\mathrm{l} = 0.5 (1 - \zeta) \sigma$; $\alpha_\mathrm{u} = 
0.5 (1 + \xi) \alpha$; $\beta_\mathrm{u} = 0.5 (1 + \zeta) \beta$;
$\gamma_\mathrm{u} = 0.5 (1 + \zeta) \gamma$ and $\sigma_\mathrm{u} =
0.5 (1 + \zeta) \sigma$.

The results of this model for cold matter are described in
more detail in Refs. \cite{MS90}-\cite{Str96} and agree rather
well with the values of the semiempirical droplet mass formula.
Interesting is also the equation of state (EOS) for cold neutron 
star matter, which is characterized by the generalized 
$\beta$-equilibrium between nucleons and leptons, i.e. electrons
and muons (for more details, see Ref. \cite{Str97}).
One has to obbey in this case the additional constraint for the 
chemical potentials, i.e.:
%%%%%%%%%%%%%%%%%%%
\begin{equation}
\mu_\mathrm{p} = \mu_\mathrm{n} - \mu_\mathrm{e};
\hspace{1cm} \mu_\mathrm{e} = \mu_\mathrm{\mu} ,
\end{equation}
%%%%%%%%%%%%%%%%%%%
and charge neutrality between the protons and the leptons:
%%%%%%%%%%%%%%%%%%%
\begin{equation}
\rho_\mathrm{p} = \rho_\mathrm{e^-} + \rho_\mathrm{\mu^-}.
\end{equation}
%%%%%%%%%%%%%%%%%%%
The EOS for this case is contained in Figs.~1 and 2.

%%%%%%%%%%%%%%%%%%%%%%%%%%%%%%%%%%%%%%%%%%%%%%%%%%%%%%%%%%%%%%%%%%%%%%%%%
\section{Thomas-Fermi model at $T > 0$} \label{sec3}

The generalization of the Thomas-Fermi model for finite temperatures
is formally more or less staightforward, since one has mainly to 
incorporate the Fermi distribution functions into
the scheme for zero temperature. 
One obtains then for the energy per nucleon:
%%%%%%%%%%%%%%%%%%%%%%%%%%
\begin{equation}
   u = \frac{2}{\rho (2\pi\hbar) ^3}\sum_{\tau} \int \limits_{-\infty}
       ^{+\infty} d^{3}p_{1}\Bigl(\frac{p_1^2}{2m_{\tau}} 
       + \frac{1}{2} V_{\tau}(p_1) \Bigr)f_{\tau}(p_{1}) ,
\end{equation}
%%%%%%%%%%%%%%%%%%%%%%%%%%
with the temperature dependent single particle potential:
%%%%%%%%%%%%%%%%%%%%%%%%%%
\begin{eqnarray}
   V_{\tau}(p_1) & = & -\frac{2}{(2\pi\hbar)^3} \frac{2T_{\mathrm{F}, \tau}}{\rho_0}
                   \nonumber\\
                   & &
                   \times \int \limits_{-\infty}^{+\infty} d^{3}p_{2}
                   \left( \alpha_{l} 
                   - \beta_{l}\left( \frac{p_{12}}{p_\mathrm{F}} \right) ^2 
                   + \gamma_{l} \frac{p_\mathrm{F}}{|p_{12}|} 
                   - \sigma_{l} \left( \frac{2 \overline{\rho}}{\rho_{0}} 
                   \right) ^{\frac{2}{3}} \right) f_{\tau}(p_{2}) 
                   \nonumber\\
                    & & - \frac{2}{(2\pi\hbar)^3} \frac{2T_\mathrm{F}}{\rho_0}
                   \nonumber\\
                   & &
                   \times
                   \int \limits_{-\infty}^{+\infty} d^{3}p_{2}
                   \left( \alpha_{u} 
                   - \beta_{u}\left( \frac{p_{12}}{p_\mathrm{F}} \right) ^2 
                   + \gamma_{u} \frac{p_\mathrm{F}}{|p_{12}|} 
                   - \sigma_{u} \left( \frac{2 \overline{\rho}}{\rho_{0}} 
                   \right) ^{\frac{2}{3}} \right) f_{-\tau}(p_{2}) ,
                   \nonumber\\
                    & & 
\end{eqnarray}
%%%%%%%%%%%%%%%%%%%%%%%%%%
and the density is now given by:
%%%%%%%%%%%%%%%%%%%%%%%%%%
\begin{equation}
   \rho = \frac{2}{(2\pi\hbar) ^3}\sum_{\tau} \int \limits_{-\infty}
          ^{+\infty} d^{3}p_{1} f_{\tau}(p_1) .
\end{equation}
%%%%%%%%%%%%%%%%%%%%%%%%%%
The Fermi distribution function, $f_{\tau}$, is defined as 
($k_\mathrm{B} = 1$):
%%%%%%%%%%%%%%%%%%%%%%%%%%
\begin{equation}
   f_{\tau}(p_{1}) = \left( 1 + \exp \left(\frac{1}{T}(\epsilon_{\tau}(p_{1}) 
                                     -\mu'_{\tau}))\right) \right)^{-1} ,
\label{fermi}
\end{equation}
%%%%%%%%%%%%%%%%%%%%%%%%%%
where $\epsilon_{\tau}$ denotes the single-particle energy:
%%%%%%%%%%%%%%%%%%%%%%%%%%
\begin{equation}
   \epsilon_{\tau}(p_{1}) = \frac{p_{1}^{2}}{2m_{\tau}} 
                            + V_{\tau}(p_{1}) .
\end{equation}
%%%%%%%%%%%%%%%%%%%%%%%%%%
With the single particle energy one can rewrite the energy per nucleon as:
%%%%%%%%%%%%%%%%%%%%%%%%%%
\begin{equation}
   u = \frac{1}{\pi^{2}\hbar^{3}\rho}
       \sum_{\tau}
       \int \limits_{0}^{+\infty}dp_{1} p_{1}^{2} \frac{1}{2}
       \left( \frac{p_{1}^{2}}{2m_{\tau}} + \epsilon_{\tau}(p_{1}) \right)
        f_{\tau}(p_{1}) .
\end{equation}
%%%%%%%%%%%%%%%%%%%%%%%%%%
Due to the momentum dependence of the single particle potential
one can also introduce an effective mass via:
%%%%%%%%%%%%%%%%%%%%%%%%%%
\begin{equation}
  m^* = \left( \frac{1}{p} \frac{\partial \epsilon(p)}{\partial p}
        \right)^{-1}_{p = 0} .
\end{equation}
%%%%%%%%%%%%%%%%%%%%%%%%%%
For finite temperatures one needs the free energy in order to obtain
the pressure and the chemical potentials.
The free energy per baryon is given by:
%%%%%%%%%%%%%%%%%%%%%%%%%%
\begin{equation}
  f = u - T s ,
\end{equation}
%%%%%%%%%%%%%%%%%%%%%%%%%%
with the entropy per baryon, $s$:
%%%%%%%%%%%%%%%%%%%%%%%%%%
\begin{equation}
  s = -\frac{2}{\rho(2\pi\hbar)^{3}}\sum_{\tau} \int \limits_{-\infty}
       ^{+\infty} d^{3}p_{1}\Bigl(f_{\tau}(p_{1}) \ln f_{\tau}(p_{1}) 
       + (1 - f_{\tau}(p_{1})) \ln (1 - f_{\tau}(p_{1}))\Bigr) .
  \label{tendsa}
\end{equation}
%%%%%%%%%%%%%%%%%%%%%%%%%%
The expressions for the chemical potentials and the pressure are:
%%%%%%%%%%%%%%%%%%%%%%%%%%
\begin{equation}
  \mu_{\tau} =\left( \rho \left(\frac{\partial}{\partial\rho_{\tau}}
  \right)_{\rho_{-\tau},T} + 1 \right) f 
\end{equation}
%%%%%%%%%%%%%%%%%%%%%%%%%%
($\mu'_{\tau}$ in Eq. \ref{fermi} is different to $\mu_{\tau}$,
since the interaction is density dependent, see Appendix A of
Ref. \cite{MS90}) 
and
%%%%%%%%%%%%%%%%%%%%%%%%%%
\begin{equation}
  P = \rho \sum_{\tau} \rho_{\tau} \left( \frac{\partial f}{\partial \rho_{\tau}}
  \right)_{\rho_{-\tau},T} .
\end{equation}
%%%%%%%%%%%%%%%%%%%%%%%%%%
More explicit expressions are given in Ref. \cite{Str96}.

%%%%%%%%%%%%%%%%%%%%%%%%%%%%%%%%%%%%%%%%%%%%%%%%%%%%%%%%%%%%%%%%%%%%%%%%%
\section{Results} \label{sec4}

First we show in Figs.~1 and 2 for the sake of completeness
the EOS for cold nuclear matter with different asymmetries
[$\delta := (\rho_\mathrm{n} - \rho_\mathrm{p}) / \rho$] and 
neutron star matter.
In Figs.~3 and 4 the EOS for symmetric nuclear
matter ($\delta = 0$) and pure neutron matter ($\delta =1$) is
displayed for different temperatures.
For small densities, below $\rho = 0.0165$ fm$^{-3}$, one obtains as
expected, the behaviour of a free Fermi gas with a linear
temperature dependence, since the nucleon-nucleon force has a small
range.
For increasing density the EOS exhibits a quadratic temperature
dependence, since in the temperature range smaller than the
Fermi temperature a quadratic temperature dependence
$u(T) = u(T=0) + a_\mathrm{v} T^2$ holds (see Refs. \cite{Str96, KWH74}).
The chemical potential $\mu$, the pressure, the free energy per
baryon and the entropy per baryon for symmetric
matter are exhibited in Figs.~5, 6, 7 and 8 for different temperatures.
The values of the entropy per baryon are in good agreement with
experimental results (see Ref. \cite{Jac83}), this is important,
since the entropy per baryon is also a nessesary ingredient for 
the calculation
of protoneutron stars (see e.g. Refs. \cite{Pra97, Str98}).

Due to the non-monotonic behaviour of the isotherms one can infer that
these systems undergo a liquid-gas phase transition.
Such a transition, which may occur in warm matter produced in
heavy-ion collisions, is one of the most interesting problems in
modern nuclear physics.
Several groups are investigating this problem by studying 
multifragmentation, where the coexistence of nuclear fragments
and free nucleons could be considered as a signal for the transition
phase (see, for instance, Refs. \cite{Gil94, Lyn97}).
The critical temperature can be obtained by a Maxwell construction.
The situation can be either viewed in the $P$-$\rho$-diagram or
in the $\mu$-$P$-diagram, respectively.
Both views are given for symmetric nuclear matter in
Figs.~9 and 10, respectively.
Gibbs condition of phase equilibrium demands that pressure, 
temperature and chemical potential are equal in the two phases.

As can be seen from the EOSs the nuclear matter incompressibility
decreases with increasing temperature.
As a consequence one should exspect a limiting temperature,
at wich a self-bound system can still exist in hydrostatic
equilibrium ($P=0$).
Beyond this temperature, the so-called flash temperature 
$T_\mathrm{fl}$, the unbound system starts expanding \cite{Sto84}.
The isothermal incompressibility is defined as:
%%%%%%%%%%%%%%%%%%%%%%%%%%
\begin{equation}
  K_T = 9 \left( \rho^2 \frac{\partial^2 f}{\partial \rho^2}
        + 2 \rho \frac{\partial f}{\partial \rho} \right)_T ,
\end{equation}
%%%%%%%%%%%%%%%%%%%%%%%%%%
and is zero at the point where the minimum pressure reaches the 
zero pressure line (here at a density of $\rho = 0.0973$ fm$^{-3}$).
We obtain for the flash temperature the value 15.61 MeV which is,
as expected, smaller than the critical temperature $T_\mathrm{c}$,
because at $T_\mathrm{c}$ the pressure is already positive
(see Fig.~6).
If one increases the asymmetry of the system the critical temperature
and the critical density decrease.
These dependencies are depicted in Figs.~11 and 12.
For asymmetries larger than 0.853 the liquid-gas phase coexistence
disappears in our model.

%%%%%%%%%%%%%%%%%%%%%%%%%%%%%%%%%%%%%%%%%%%%%%%%%%%%%%%%%%%%%%%%%%%%%%%%%
\section{Conclusions} \label{sec5}

We have calculated the EOS for cold and warm nuclear matter for
different asymmetries using the approach of Myers and \'Swi\c atecki.
Almost all non-relativistic calculations give critical temperatures
in the range of 14 - 22 MeV and critical densities
of approximately $1/3$ of the saturation density \cite{BGLBF95}, 
which is in accordance with our results.
Investigations for finite temperatures are relatively scare and
the model seems to provide a simple theoretical tool for 
investigating the nuclear EOS in such ranges, which seems
to be of greater interest, since it becomes now possible by
means of the heavy-ion physics to study the behaviour of matter
different from the saturation point, which is an essential
ingredient in the theory of neutron stars and supernova 
explosions and also in the interpolation of heavy-ion
collisions \cite{LKB97}.

In our opinion, it seems worthwhile, due to its simplizity
and exellent agreement with nuclear data, to use this model
with possible improvements in the further non-relativistic 
investigations of the EOS.
Also the properties of neutron stars seem to be in accordance 
with this approach \cite{Str97}.

%%%%%%%%%%%%%%%%%%%%%%%%%%%%%%%%%%%%%%%%%%%%%%%%%%%%%%%%%%%%%%%%%%%%%%%%%
\bigskip
\section*{Acknowledgments:} The authors are indebted to 
W. D. Myers, W. J. \'Swi\c atecki,
W. Stocker and Ch. Schaab for valuable discussions.
One of us, K. S., gratefully acknowleges the Bavarian State for financial 
support.
%%%%%%%%%%%%%%%%%%%%%%%%%%%%%%%%%%%%%%%%%%%%%%%%%%%%%%%%%%%%%%%%%%%%%%%%%

\clearpage

%%%%%%%%%%%%%%%%%%%%%%%%%%%%%%%%%%%%%%%%%%%%%%%%%%%%%%%%%%%%%%%%%%%%%%%%%
\section*{Table captions}

\begin{description}
\item[TABLE~I:] Parametrization of the interaction \cite{MS90}.
\end{description}

%%%%%%%%%%%%%%%%%%%%%%%%%%%%%%%%%%%%%%%%%%%%%%%%%%%%%%%%%%%%%%%%%%%%%%%%%
\clearpage

%%%%%%%%%%%%%%%%%%%%%%%%%%%%%%%%%%%%%%%%%%%%%%%%%%%%%%%%%%%%%%%%%%%%%%%%%
\begin{table}
\begin{center}
TABLE I \\[1cm]
\begin{tabular}{ c c c c }
   \hline
      &    &    &   \\
   Parameter $x$ & Value & Value $x_{l}$ & Value $x_{u}$ \\
      &    &    &   \\
   \hline
   &    &    &   \\
   a & 0.59542 fm & - & - \\
   $\alpha$ & 3.60928 & 1.01054 & 2.59874 \\
   $\beta$ & 0.37597 & 0.07561 & 0.30036 \\
   $\gamma$ & 0.21329 & 0.04289 & 0.17040 \\
   $\sigma$ & 1.33677 & 0.26884 & 1.06793 \\
   B & 1.48995 & 0.29964 & 1.19030 \\
   $\xi$ & 0.44003 & - & - \\
   $\zeta$ & 0.59778 & - & - \\
   &    &    &   \\
   \hline
\end{tabular}
\end{center}
\end{table}

%%%%%%%%%%%%%%%%%%%%%%%%%%%%%%%%%%%%%%%%%%%%%%%%%%%%%%%%%%%%%%%%%%%%%%%%%
\clearpage
%%%%%%%%%%%%%%%%%%%%%%%%%%%%%%%%%%%%%%%%%%%%%%%%%%%%%%%%%%%%%%%%%%%%%%%%%
\section*{Figure captions}

\begin{description}

\item[Fig.~1.] Energy per nucleon versus density for nuclear matter
for different asymmetries and neutron star matter 
in $\beta$-equilibrium.
The nuclear matter parameter are:
$\rho_0 = 0.16545$ fm$^{-3}$; $u = 16.527$ MeV; 
incompressibility $K = 301.27$ MeV; asymmetry parameter
$J = 31.375$ MeV.
The maximum star mass for a non-rotating neutron star is
approximately 2.0 solar masses and the star radii are between
10 and 12 km.
  
\item[Fig.~2.] Pressure as function of density for cold nuclear matter 
for different asymmetries and neutron star matter.

\item[Fig.~3.] Energy per nucleon for symmetric nuclear
matter ($\delta = 0$) versus density at different temperatures.

\item[Fig.~4.] Energy per nucleon neutron
matter ($\delta = 1$) versus density at different temperatures.

\item[Fig.~5.] Chemical potential as function of density
for symmetric nuclear matter at different temperatures.

\item[Fig.~6.] Pressure versus density for symmetric nuclear matter 
at different temperatures.

\item[Fig.~7.] Free energy per baryon versus density for symmetric 
nuclear matter at different temperatures.

\item[Fig.~8.] Entropy per baryon versus density for symmetric 
nuclear matter at different temperatures.
The entropy behaviour agrees with the experimental situation
(cf. with Fig. 6 of Ref. \cite{DTS92}, see also Ref. \cite{Jac83}).

\item[Fig.~9.] Liquid-gas phase transition in the $P$-$\rho$
diagram for symmetric matter:
Shown are the isoterms for $T$ = 0, 10, 15.61, 20.8 and 30 MeV.
Above the dashed-dotted curve the matter exists only in one
phase, either as gas (g) between the dashed-dotted line and the
isotherm with the critical temperature $T_\mathrm{c}$ = 20.8 MeV
or as liquid (f) above $T_\mathrm{c}$ = 20.8 MeV.
Between the dashed-dotted curve and the dashed curve one finds
metastable phases, again either as a gas (g') or a liquid (f').
Below the dashed curve the state is not stable. 
Shown is also the pressure at the flash temperature (15.61 MeV).

\item[Fig.~10.] Chemical potential as function of the pressure
for symmetric nuclear matter.
The crossing point on each isotherm is the condition of
phase equilibrium.

\item[Fig.~11.] Asymmetry dependence of the critical
temperature $T_\mathrm{c}$.

\item[Fig.~12.] Asymmetry dependence of the critical
density $\rho_\mathrm{c}$.

\end{description}

\clearpage
%%%%%%%%%%%%%%%%%%%%%%%%%%%%%%%%%%%%%%%%%%%%%%%%%%%%%%%%%%%%%%%%%%%%%%%%%
%%%%%%%%%%%%%%%%%%%%%%%%%%%%%
\begin{figure}[tbp] 
\begin{center}  
\leavevmode 
FIGURE 1\\[0.5cm]
\epsfig{figure=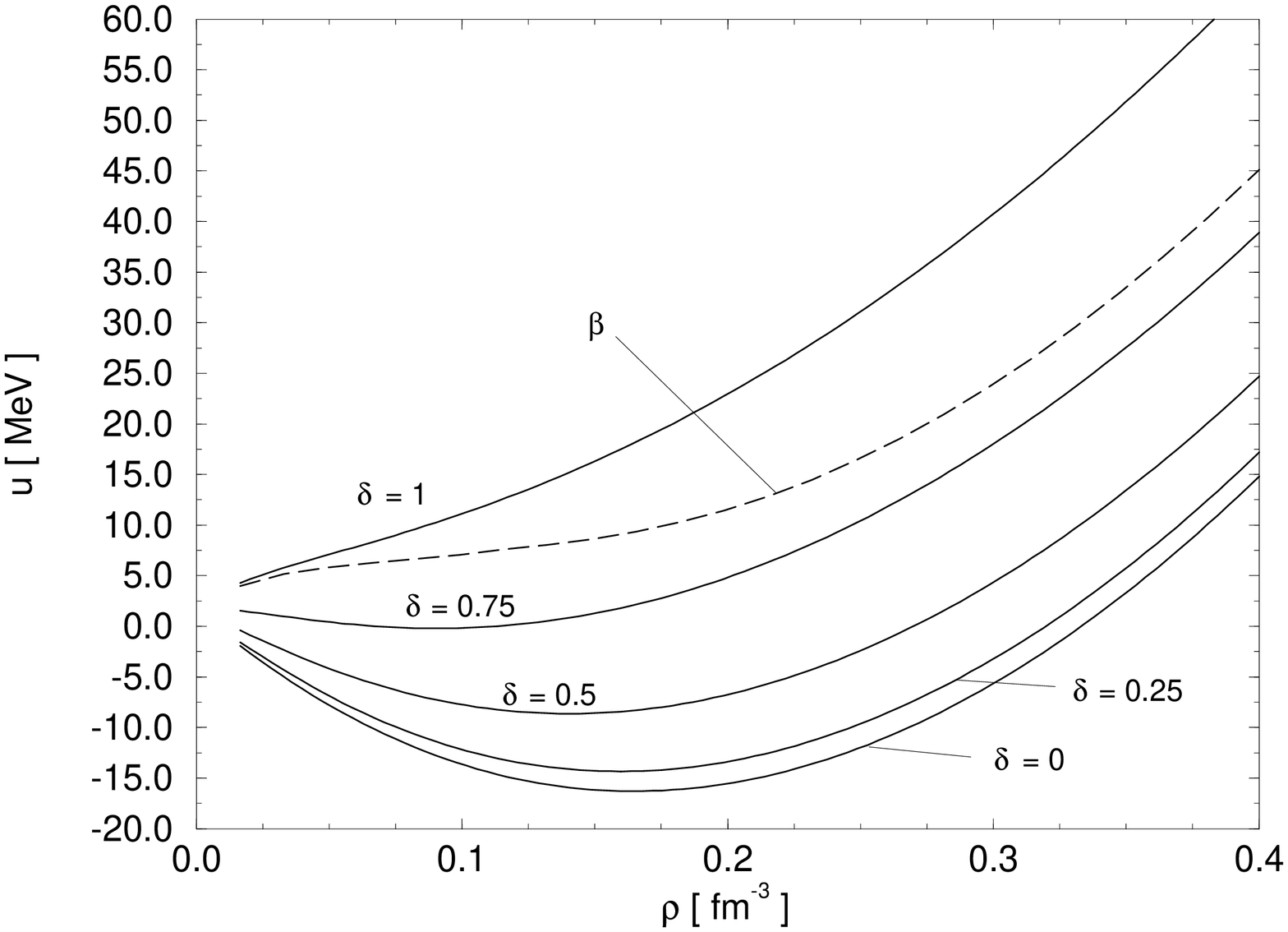,width=70mm,clip=,angle=-90,
           bbllx=95,bblly=32,bburx=560,bbury=675}
\end{center}
\end{figure}
%%%%%%%%%%%%%%%%%%%%%%%%%%%%%
\begin{figure}[tbp] 
\begin{center}  
\leavevmode 
FIGURE 2\\[0.5cm]
\epsfig{figure=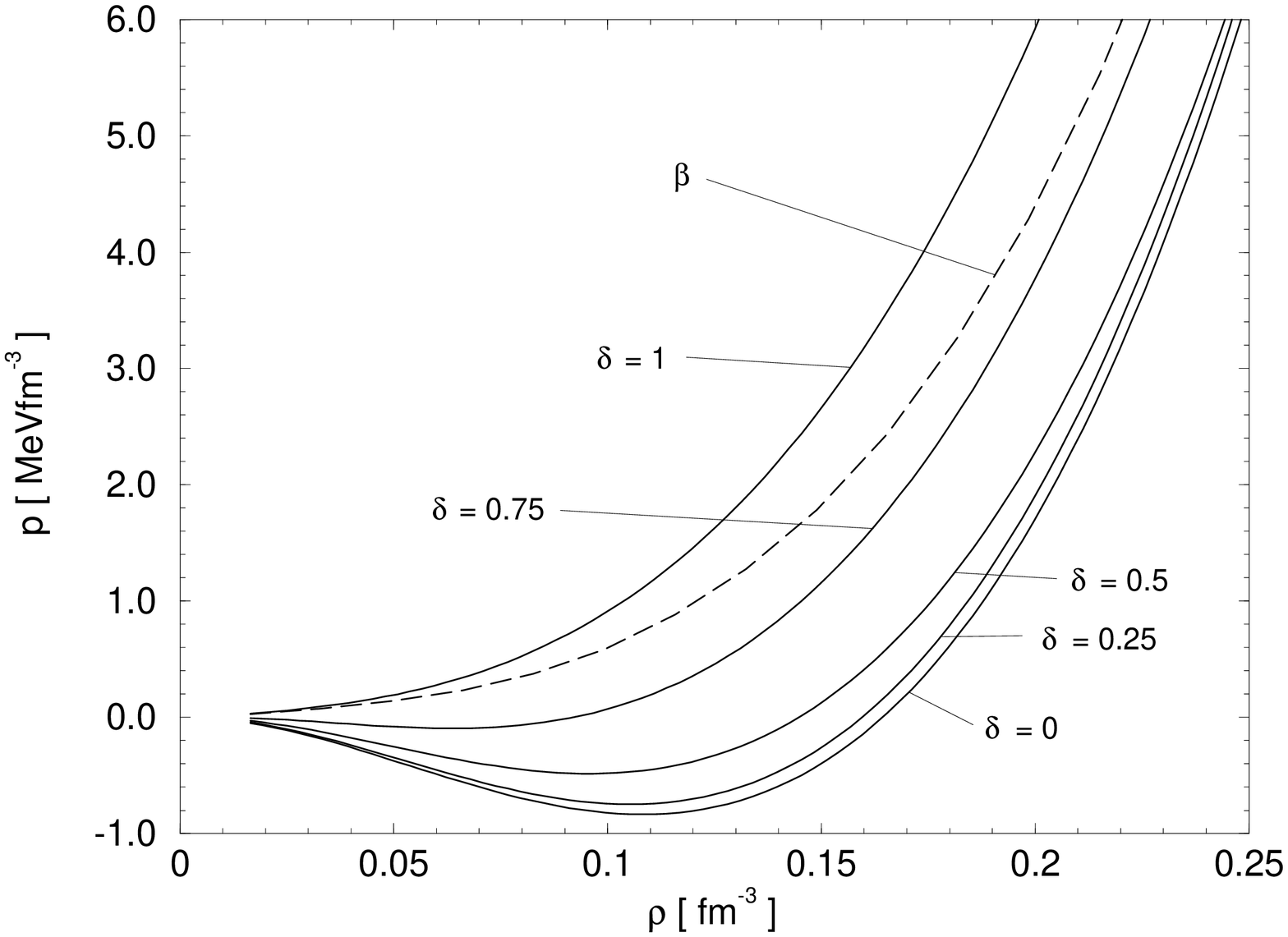,width=70mm,clip=,angle=-90,
           bbllx=95,bblly=42,bburx=560,bbury=680}
\end{center}
\end{figure}
%%%%%%%%%%%%%%%%%%%%%%%%%%%%%
\begin{figure}[tbp] 
\begin{center}  
\leavevmode 
FIGURE 3\\[0.5cm]
\epsfig{figure=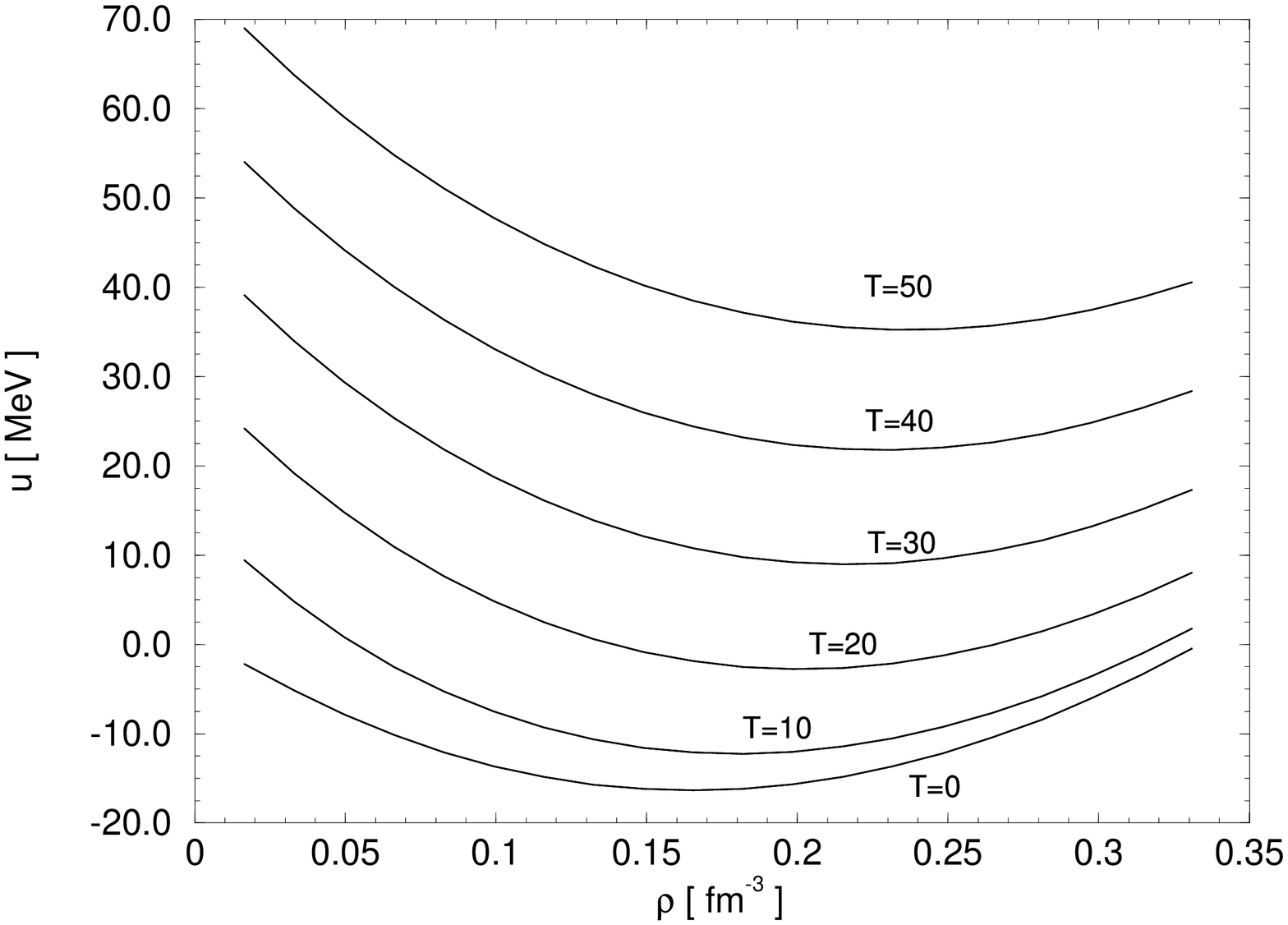,width=70mm,clip=,angle=-90,
           bbllx=95,bblly=32,bburx=560,bbury=680}
\end{center}
\end{figure}
%%%%%%%%%%%%%%%%%%%%%%%%%%%%%
\begin{figure}[tbp] 
\begin{center}  
\leavevmode 
FIGURE 4\\[0.5cm]
\epsfig{figure=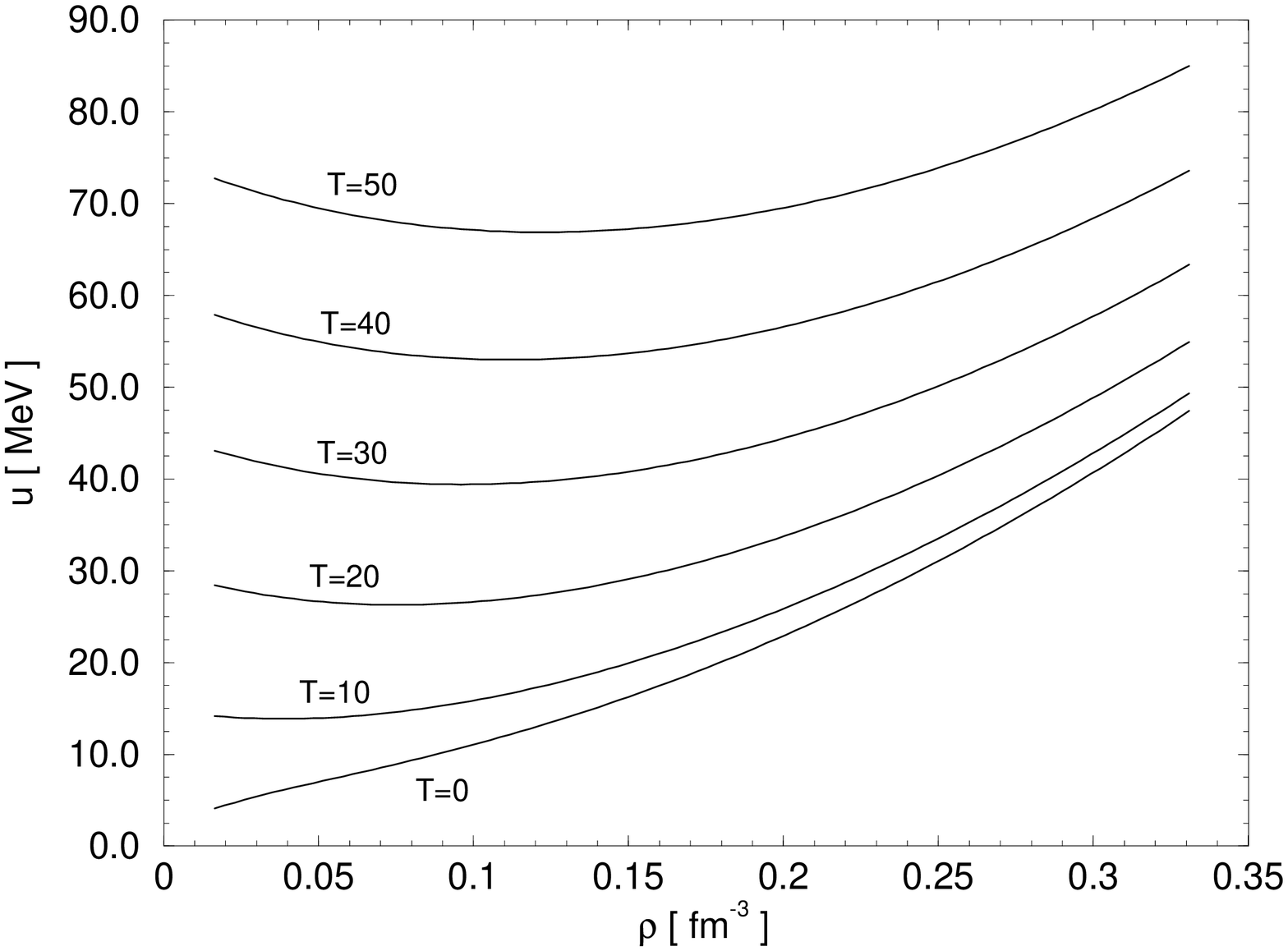,width=70mm,clip=,angle=-90,
           bbllx=95,bblly=50,bburx=560,bbury=680}
\end{center}
\end{figure}
%%%%%%%%%%%%%%%%%%%%%%%%%%%%%
\begin{figure}[tbp] 
\begin{center}  
\leavevmode 
FIGURE 5\\[0.5cm]
\epsfig{figure=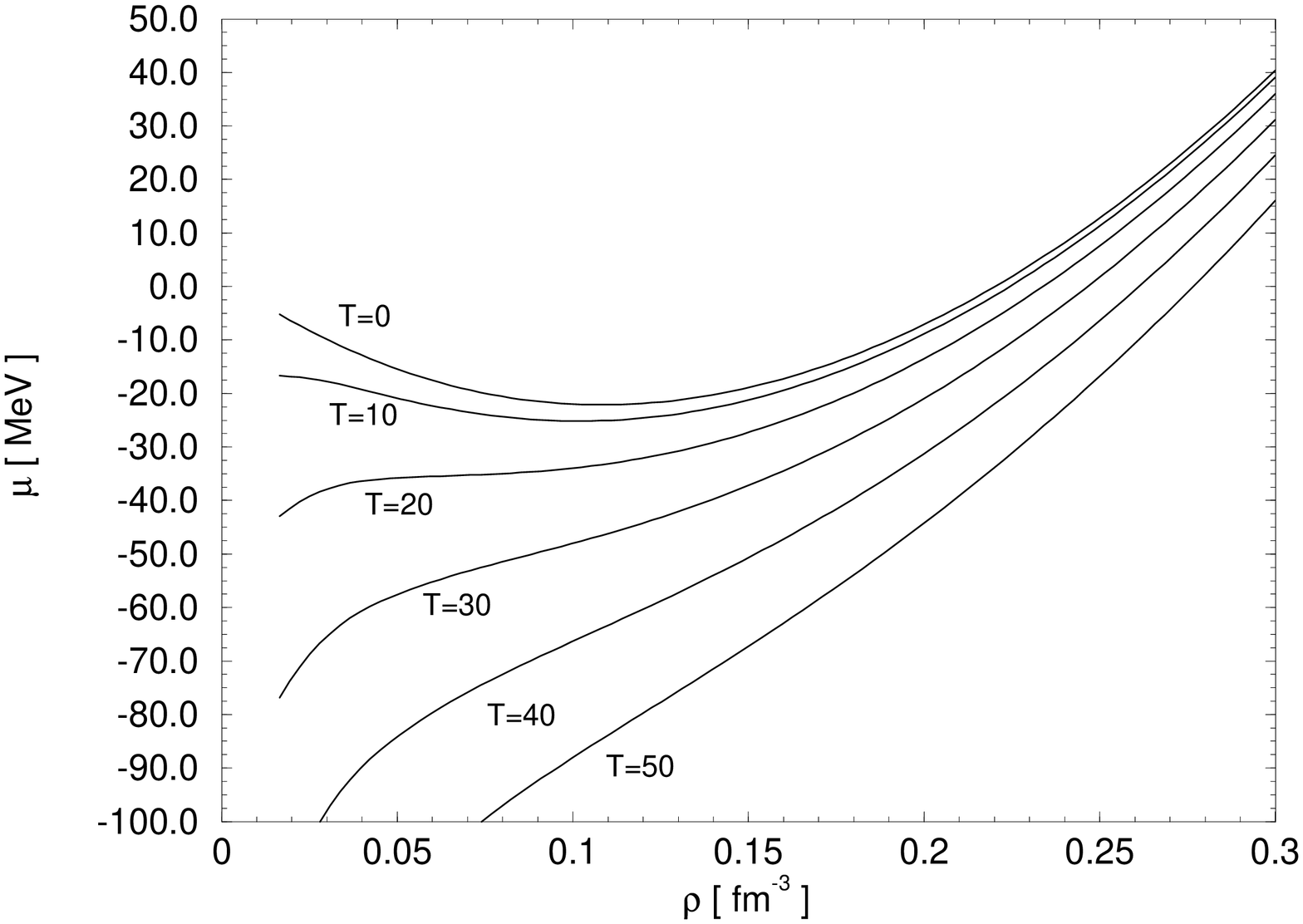,width=70mm,clip=,angle=-90,
           bbllx=95,bblly=20,bburx=560,bbury=675}
\end{center}
\end{figure}
%%%%%%%%%%%%%%%%%%%%%%%%%%%%%
\begin{figure}[tbp] 
\begin{center}  
\leavevmode 
FIGURE 6\\[0.5cm]
\epsfig{figure=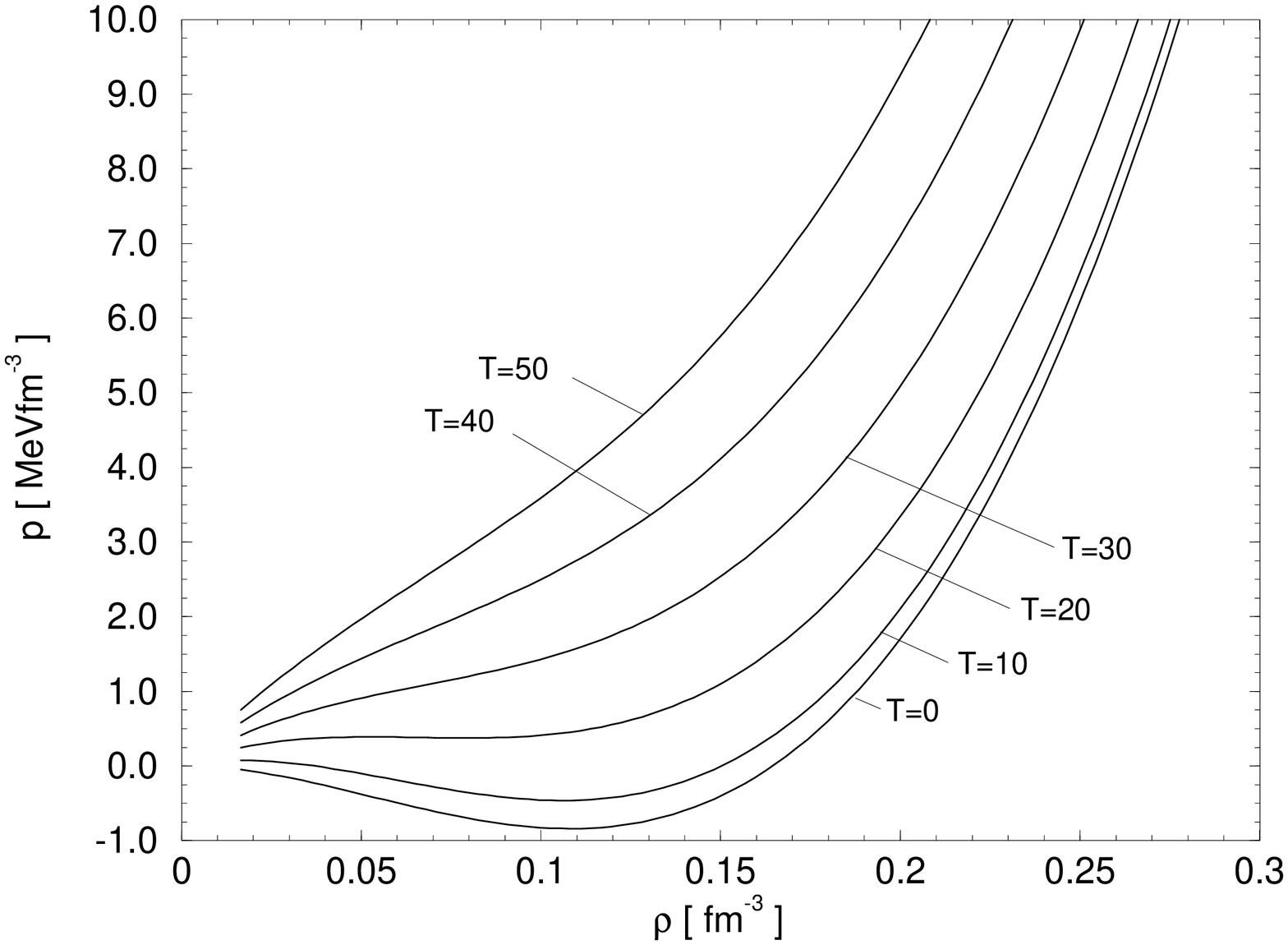,width=70mm,clip=,angle=-90,
           bbllx=95,bblly=42,bburx=560,bbury=675}
\end{center}
\end{figure}
%%%%%%%%%%%%%%%%%%%%%%%%%%%%%
\begin{figure}[tbp] 
\begin{center}  
\leavevmode 
FIGURE 7\\[0.5cm]
\epsfig{figure=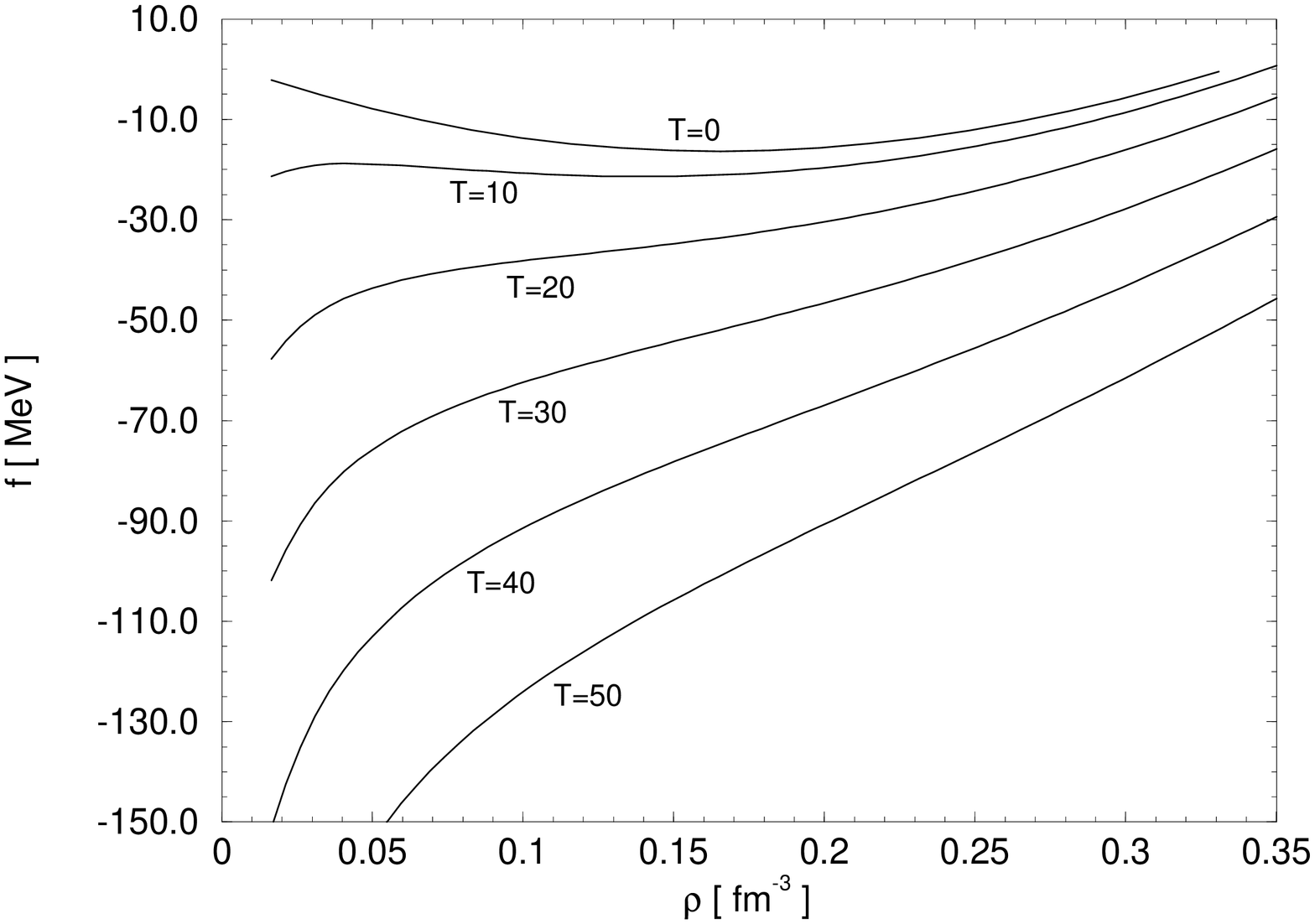,width=70mm,clip=,angle=-90,
           bbllx=95,bblly=20,bburx=560,bbury=680}
\end{center}
\end{figure}
%%%%%%%%%%%%%%%%%%%%%%%%%%%%%
\begin{figure}[tbp] 
\begin{center}  
\leavevmode 
FIGURE 8\\[0.5cm]
\epsfig{figure=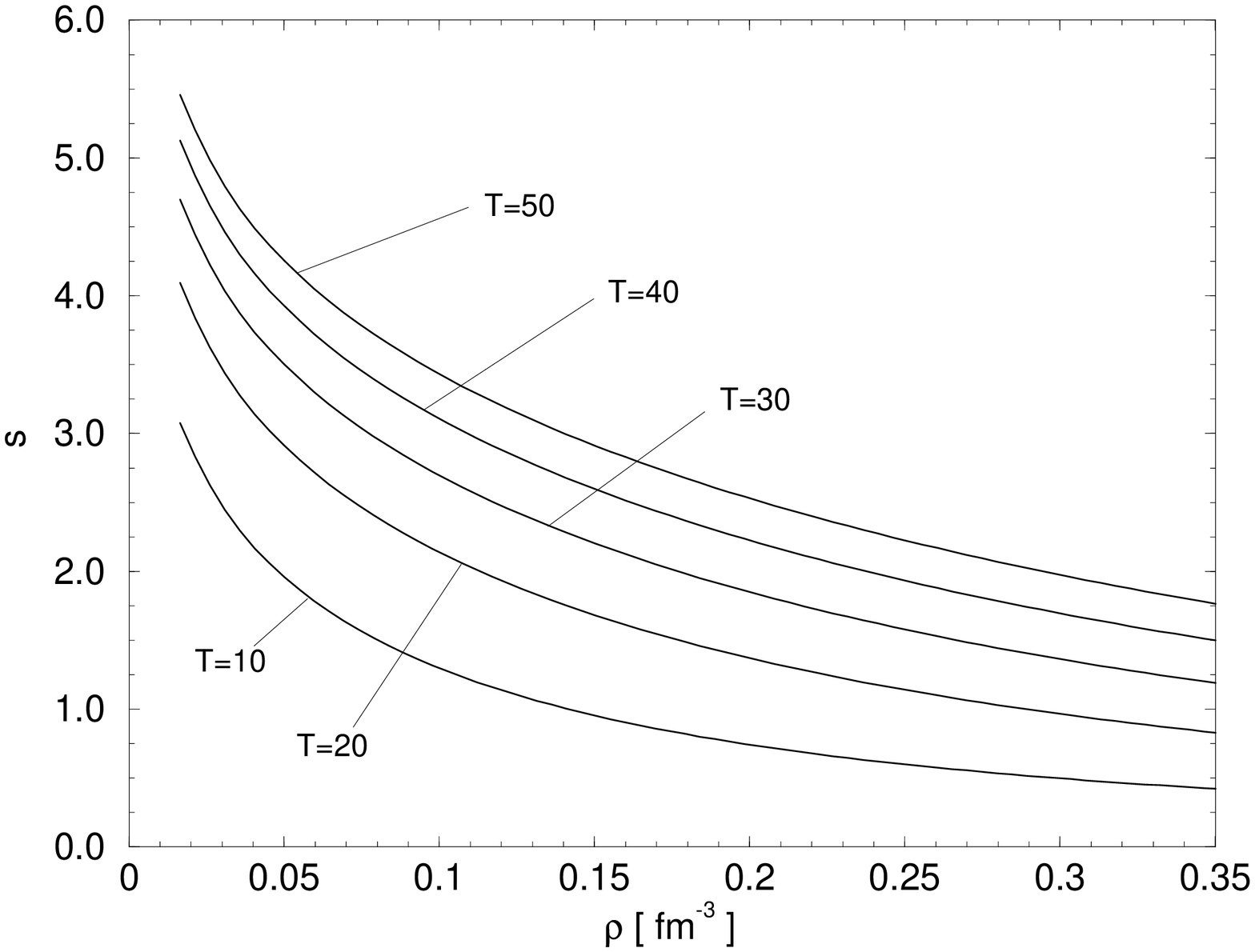,width=70mm,clip=,angle=-90,
           bbllx=95,bblly=68,bburx=560,bbury=680}
\end{center}
\end{figure}
%%%%%%%%%%%%%%%%%%%%%%%%%%%%%
\begin{figure}[tbp] 
\begin{center}  
\leavevmode 
FIGURE 9\\[0.5cm]
\epsfig{figure=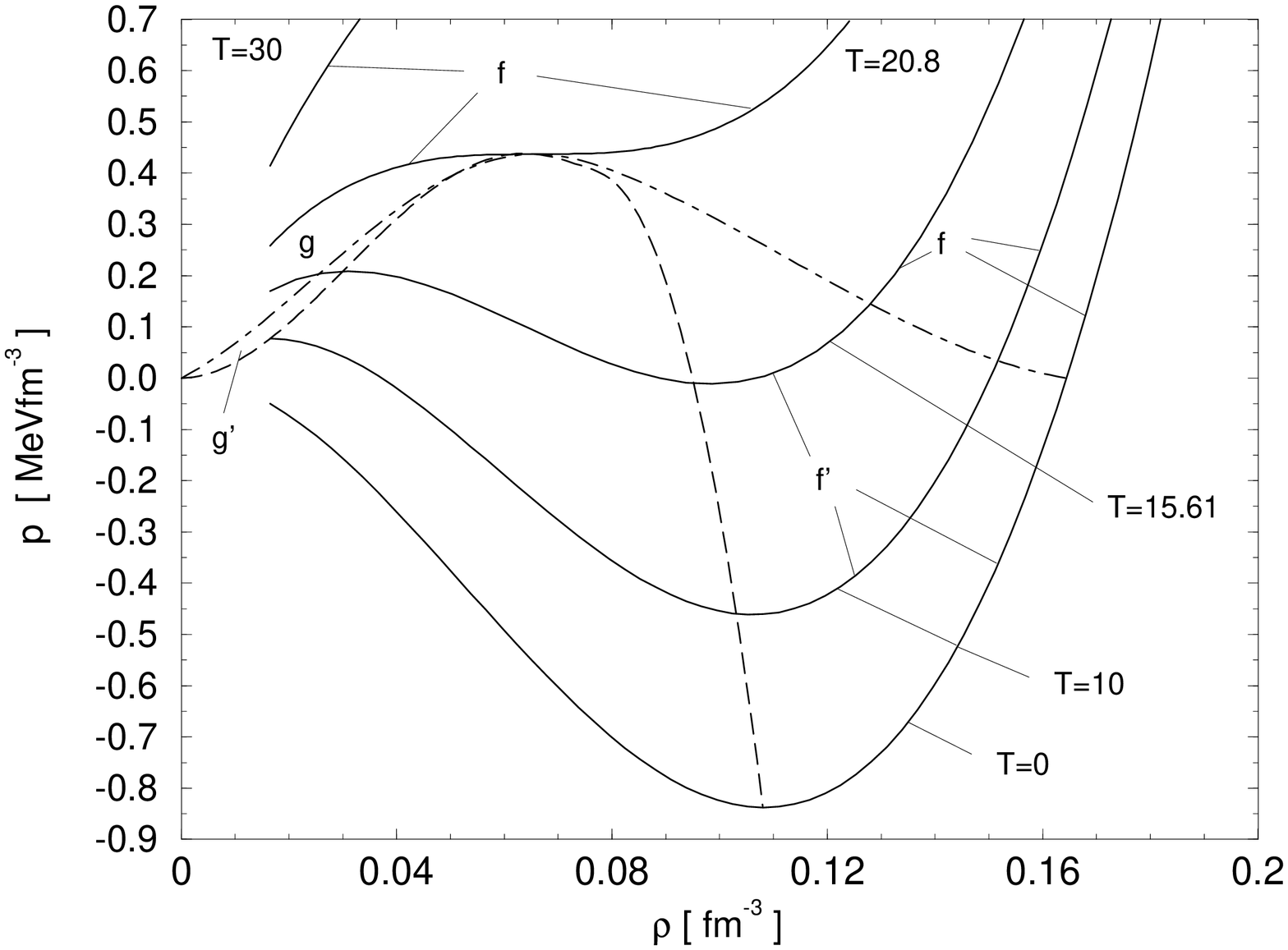,width=70mm,clip=,angle=-90,
           bbllx=95,bblly=42,bburx=562,bbury=675}
\end{center}
\end{figure}
%%%%%%%%%%%%%%%%%%%%%%%%%%%%%
\begin{figure}[tbp] 
\begin{center}  
\leavevmode 
FIGURE 10\\[0.5cm]
\epsfig{figure=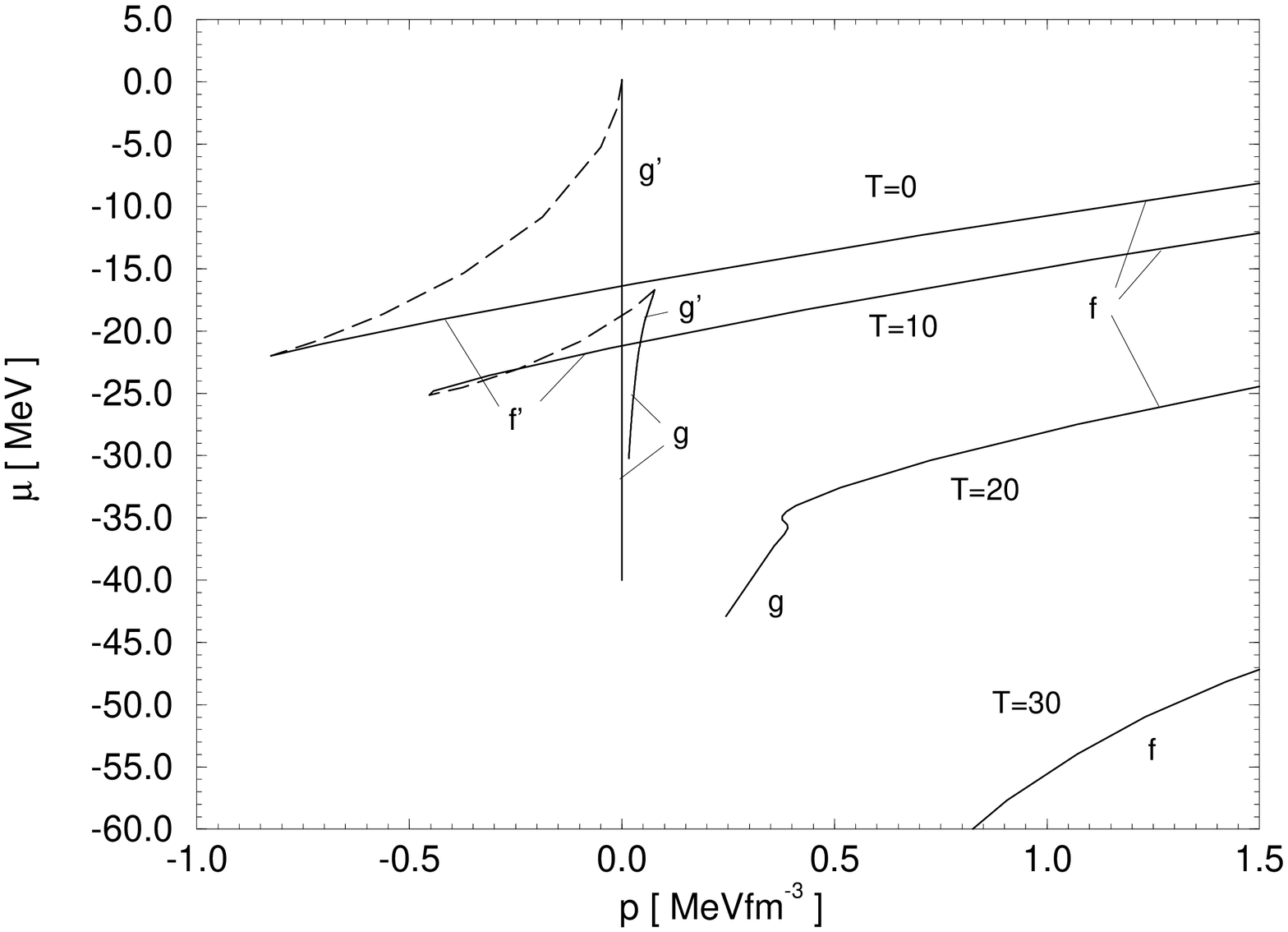,width=70mm,clip=,angle=-90,
           bbllx=95,bblly=32,bburx=560,bbury=675}
\end{center}
\end{figure}
%%%%%%%%%%%%%%%%%%%%%%%%%%%%%
\begin{figure}[tbp] 
\begin{center}  
\leavevmode 
FIGURE 11\\[0.5cm]
\epsfig{figure=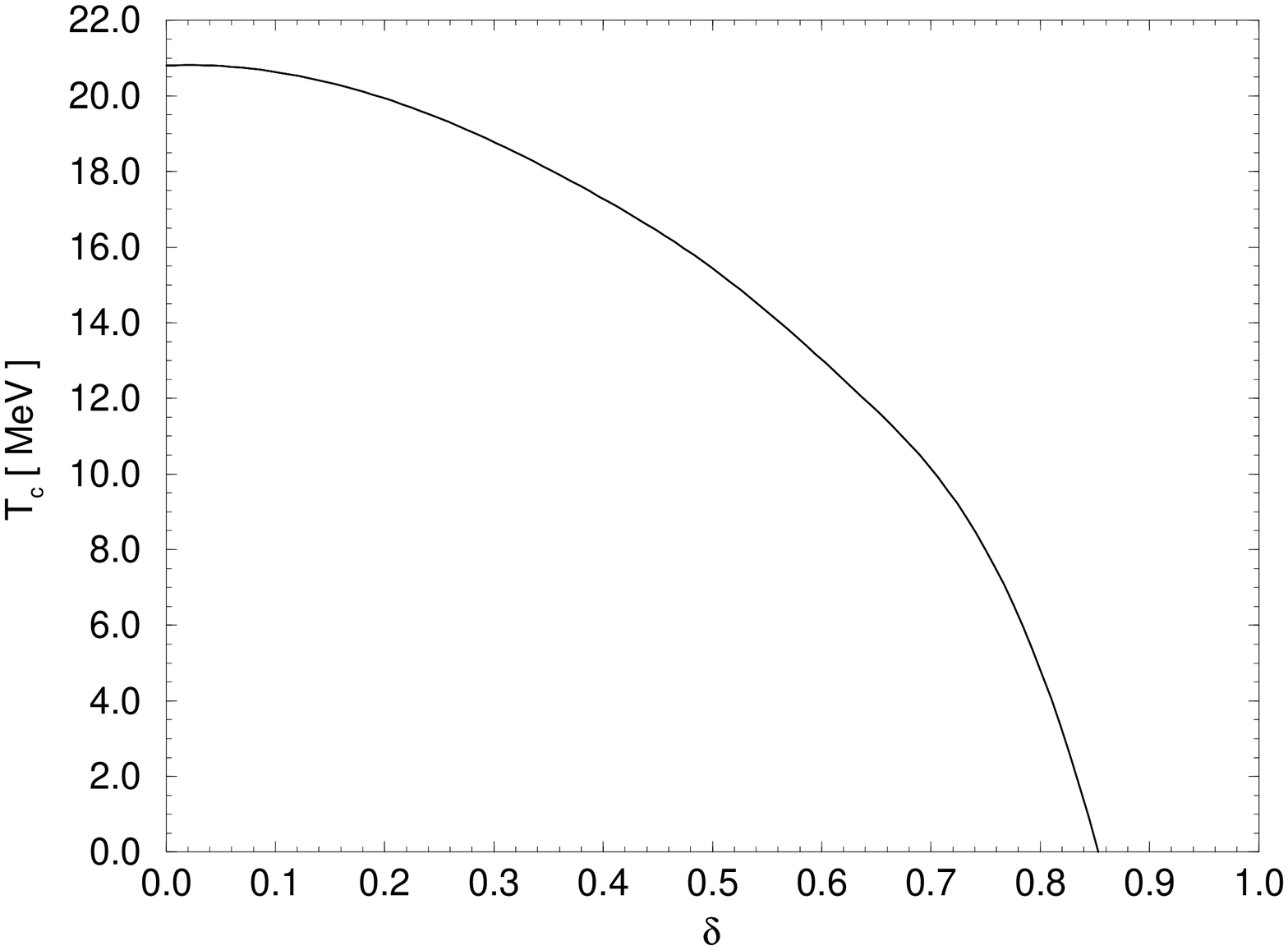,width=70mm,clip=,angle=-90,
           bbllx=95,bblly=50,bburx=555,bbury=675}
\end{center}
\end{figure}
%%%%%%%%%%%%%%%%%%%%%%%%%%%%%
\begin{figure}[tbp] \label{fig10}
\begin{center}  
\leavevmode 
FIGURE 12\\[0.5cm]
\epsfig{figure=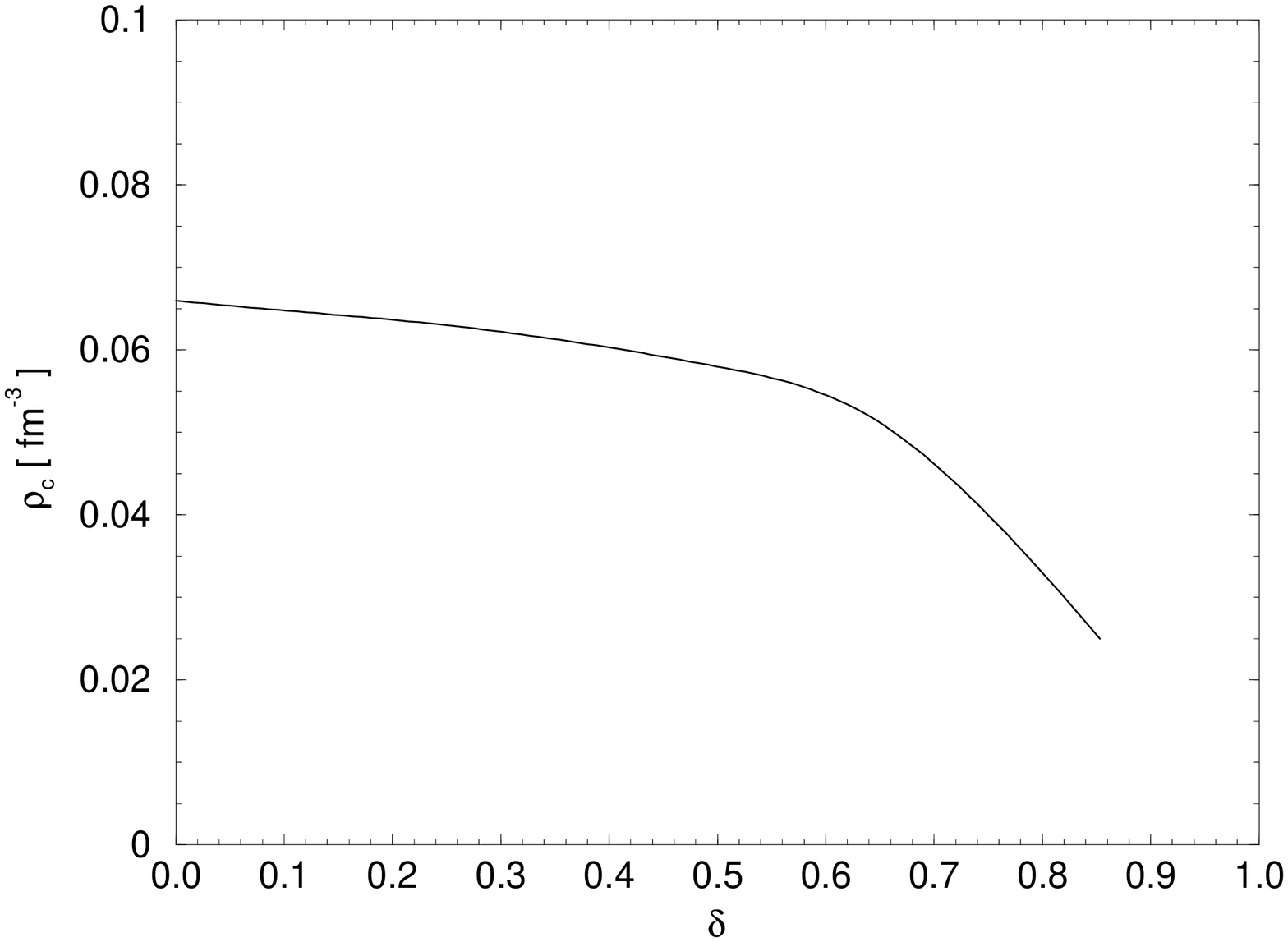,width=70mm,clip=,angle=-90,
           bbllx=95,bblly=45,bburx=555,bbury=675}
\end{center}
\end{figure}
%%%%%%%%%%%%%%%%%%%%%%%%%%%%%
%%%%%%%%%%%%%%%%%%%%%%%%%%%%%%%%%%%%%%%%%%%%%%%%%%%%%%%%%%%%%%%%%%%%%%%%%
\clearpage
%%%%%%%%%%%%%%%%%%%%%%%%%%%%%%%%%%%%%%%%%%%%%%%%%%%%%%%%%%%%%%%%%%%%%%%%%


\begin{thebibliography}
{99}
\bibitem{Ser92} B.~D.~Serot, Rep. Prog. Phys. {\bf 55}, 
1855 (1992), and references therein.
\bibitem{WW88} F.~Weber and M.~K.~Weigel, Z. Phys. {\bf A330},
249 (1988).
\bibitem{HWW98} H.~Huber, F.~Weber, and M.~K.~Weigel, Phys. 
Rev. C {\bf 57}, 3484 (1998), and references therein.
\bibitem{BGLBF95} M.~Baldo, G.~Giansiracuso, U.~Lombardo,
I.~Bombaci, and L.~S.~Ferreira, Nucl. Phys. {\bf A583},
599c (1995), and references therein.
\bibitem{PRL95} C.~J.~Pethick, D.~G.~Ravenhall, and C.~P.~Lorentz
{\bf A584}, 675 (1995), and references therein.
\bibitem{DTS92} C.~Das, R.~K.~Tripathi, and R.~Sahu, Phys. Rev. C
{\bf 45}, 2217 (1992).
\bibitem{MS90} W.~D.~Myers and W.~J.~\'Swi\c atecki, Ann. of Phys. {\bf 204}, 401
(1990).
\bibitem{MS91} W.~D.~Myers and W.~J.~\'Swi\c atecki, Ann. of Phys. {\bf 211}, 292 
(1991).
\bibitem{MS96} W.~D.~Myers and W.~J.~\'Swi\c atecki, Nucl. Phys. {\bf A601}, 141
(1996).
\bibitem{MS98} W.~D.~Myers and W.~J.~\'Swi\c atecki, Phys. Rev. C {\bf 57}, 3020
(1998).
\bibitem{Str96} K.~Strobel, Master's thesis, Universit\"at M\"unchen,
1996, unpublished.
\bibitem{Tho26} L.~H.~Thomas, Proc. Cambridge Phil. Soc. {\bf 23},
542 (1926).
\bibitem{Fer28} E.~Fermi, Z. Phys. {\bf 48}, 73 (1928).
\bibitem{MS69} W.~D.~Myers and W.~J.~\'Swi\c atecki, Ann. of Phys. {\bf 55},
395 (1969).
\bibitem{SB61} R.~G.~Seyler and C.~H.~Blanchard, Phys. Rev. {\bf 124},
227 (1961); {\bf 131}, 355 (1963).
\bibitem{Str97} K.~Strobel, F.~Weber, Ch.~Schaab, and M.~K.~Weigel, 
Int. J. Mod. Phys. E {\bf 6}, 669 (1997).
\bibitem{KWH74} W.~A.~K\"upper, G.~Wegmann, and E.~R.~Hilf, 
Ann. of Phys. {\bf 88}, 454 (1974).
\bibitem{Pra97} M.~Prakash, I.~Bombaci, M.~Prakash, P.~J.~Ellis,
J.~M.~Lattimer, and R.~Knorren, Phys. Rep. {\bf 280}, 1 (1997).
\bibitem{Str98} K.~Strobel, Ch.~Schaab, and M.~K.~Weigel, 
in preparation.
\bibitem{Jac83} B.~V.~Jacak et al., Phys. Rev. Lett. {\bf 51}, 
1846 (1983).
\bibitem{Gil94} M.~L.~Gilkeset et al., Phys. Rev. Lett. {\bf 73}, 
1590 (1994).
\bibitem{Lyn97} U.~Lynen, ``Multifragmentation and the Search for the 
Liquid-Gas Phase Transition in Nuclear matter'', in Proc. of the 
International Conference on Nuclear Physics at the Turn of the 
Millenium, Wilderness, South Africa, 1996, edited by 
H.~St\"ocker, A.~Gallmann, and J.~H.~Hamilton (World Scientific,
Singapore, 1997), p. 200, and references therein.
\bibitem{Sto84} W.~Stocker, Phys. Lett. {\bf 142B}, 319 (1984).
\bibitem{LKB97} Baon-An~Li, Che~Ming~Ko, and W.~Bauer,
TAMU-Nucl.-Th-97-04; J. Phys. G (to be published), 
for a review and further references.


\end{thebibliography}
\end{document}